\documentclass{article}
\usepackage{natbib}
\usepackage{setspace}
\usepackage{amssymb,amsmath,amsfonts,eurosym,geometry,ulem,graphicx,caption,color,setspace,sectsty,comment,footmisc,pdflscape,subfigure,array,hyperref,booktabs,tabularx,multirow,float,verbatim,url,lipsum}
\usepackage[section]{placeins}

\newcolumntype{L}[1]{>{\raggedright\let\newline\\arraybackslash\hspace{0pt}}m{#1}}
\newcolumntype{C}[1]{>{\centering\let\newline\\arraybackslash\hspace{0pt}}m{#1}}
\newcolumntype{R}[1]{>{\raggedleft\let\newline\\arraybackslash\hspace{0pt}}m{#1}}

\usepackage{hyperref}
\hypersetup{
    colorlinks=true,
    citecolor=blue,
    filecolor=blue,
    linkcolor=blue,
    urlcolor=blue,
  pdfstartview={FitR},
filecolor=blue
}
\begin{document}

\begin{titlepage}
\title{Sentiment Analysis of State Bank of Pakistan's Monetary Policy Documents and it's Impact on Stock Market}
\author{Aabid Karim, Heman Das Lohano}
\date{\today\\ \vspace{1em}Institute of Business Administration, Karachi}
\maketitle

\begin{abstract}
\noindent 
This research examines whether sentiments conveyed in the State Bank of Pakistan's (SBP) communications impact financial market expectations and can act as a monetary policy tool. To achieve our goal, we first use sentiment analysis techniques to quantify the tone of SBP monetary policy documents and second, we use short time window, high frequency methodology to approximate the impact of tone on stock market returns. Our results show that positive (negative) change in the tone  positively (negatively) impacts stock returns in Karachi Stock Exchange. Further extension shows that the communication of SBP still has a statistically significant impact on stock returns when controlling for different variables and monetary policy tool. Also, the communication of SBP does not have a long term constant effect on stock market. 
\\
\vspace{0in}
\end{abstract}

\setcounter{page}{0}
\thispagestyle{empty}
\end{titlepage}

\pagebreak

\doublespacing

\tableofcontents
\newpage
\section{Introduction} \label{sec:introduction}
The Pakistani stock market plays a crucial role in the country's economic development, and there is evidence suggesting a long-run relationship between the Pakistani stock market and economic growth \cite{shahbaz2008stock}. It serves as a barometer of investor confidence, facilitates capital formation, and drives industrial activity \cite{tauseef2022pakistan}. However, understanding the factors influencing its performance remains a complex challenge. This research aims to delve into 
one such potentially influential factor: the sentiments conveyed in the State Bank of Pakistan's (SBP) monetary policy documents. Central banks worldwide employ monetary policy pronouncements to guide economic expectations and influence market behavior. These pronouncements often utilize specific language styles and tones, potentially impacting investor sentiment and subsequent investment decisions \cite{blinder2008central}. Despite this potential influence, the relationship between central bank language and stock market performance remains under-researched, particularly in emerging markets like Pakistan.

\noindent
While existing studies have explored the impact of central bank communication on various factors of the economy, examined its influence on stock market movements. In the context of Pakistan, no prior research has investigated the link between Central Bank of Pakistan's communication and stock market of Pakistan. This study aims to bridge this gap. 

\noindent
This research proposes the following research question: How does the sentiments conveyed in the State Bank of Pakistan's monetary policy documents affect the performance of the KSE - 100 index?

\noindent
H1: Increased optimism or confidence expressed in SBP communications positively impacts KSE - 100 performance. 

\noindent
H2: Aggressive or hawkish stances adopted in SBP pronouncements negatively affect KSE -1 00 performance.

\noindent
This research employs a quantitative approach based on time-series data analysis. We utilize sentiment analysis techniques to extract sentiment indicators from SBP documents from 2016 to 2024. These sentiment scores  regresse against the KSE - 100 index returns, controlling for relevant macroeconomic and financial variables such as policy rate, economic uncertainty, and consumer confidence etc.  
Understanding the link between SBP language and stock market performance in Pakistan, can have significant implications for both policymakers and investors. For policymakers, it can provide valuable insights into crafting communication strategies that foster investor confidence and market stability. For investors, it can offer valuable predictive tools for navigating the market based on central bank pronouncements.

\noindent
In conclusion, this research promises to address a critical gap in our understanding of state bank communication and its impact on Pakistani stock market performance. By analyzing the relationship between SBP sentiment and KSE - 100 movement, we hope to contribute valuable insights to both academic and practical domains, thereby aiding in fostering sustainable growth and stability in the Pakistan financial system.

\section{Literature Review} \label{sec:Literature Review}
This literature review delves into the growing field of central bank communications, sentiment analysis, and their impact on financial markets. By investigating diverse studies across geographical contexts and central banks, this review aims to establish a comprehensive understanding of how central bank pronouncements influence market behavior and contribute to financial stability. This literature suggests that the most direct effects of monetary policy, along with central bank communications, are on the financial sector. To be more specific, \cite{bennani2019does} stated that changes in monetary policy are transmitted through the stock market via changes in the values of private portfolios or changes in the cost of capital. To explain this \cite{bernanke2005explains}  estimated that 25-basis-point cut in the federal funds rate target is linked with about a 1 percent increase in CRSP value-weighted index. They found that impacts of unexpected monetary policy actions on expected excess returns take for the largest part of the response of stock prices. \cite{de2008effect} explains that financial markets clearly respond to the effective communication of central banks. \cite{gurkaynak2005sensitivity} also argues that asset prices respond to the Open Market Committee (FOMC)
statements. They estimated that FOMC statements put their effects on financial market
via their influence on market expectations. \cite{ehrmann2007explaining} and \cite{reeves2007financial} concluded the same for European Central Bank and the Bank of England
respectively.

\vspace{\baselineskip}
\noindent
However, the transparency and communication strategies have been increasing of the State Bank of Pakistan but little is known how these communication affects stock market of Pakistan \cite{sohail2019understanding}, \cite{zaheer2019effectiveness}, and \cite{saad2019understanding}. The SBP releases statements and publish monetary policy documents along with annual financial reports about monetary policy and economic outlook. So, looking at the unobserved nature
of various tools of monetary policy implemented by SBP, its announcements provides news which are expected to move the assets prices. \cite{xiong2012measuring} computed an index of monetary policy based on People Bank of China documents.  In order to estimate
a monetary policy rule, the author codes People Bank of China (PBC) statements and uses this variable. He concludes that words aid deeds in understanding PBC monetary policy after finding that his rule performs better than one based solely on interest rates. But we should bear in mind that central banks have no full power in their objectives and instruments. As a result, the central bank’s ability to provide prior advice on when to make crucial monetary policy decisions is limited \cite{mcmahon2018china}. 

\vspace{\baselineskip}
\noindent
Moreover, existing literature states that sentiments conveyed in the documents of central
banks determines, how market agents develope and alter their expectations \cite{angeletos2013sentiments}, \cite{angeletos2018quantifying}, and \cite{benhabib2015sentiments}. Hence, sentiments conveyed in the monetary policy documents and speeches of the central banks are considered to be valuable information \cite{hubert2017central}. If we dive more deeply into the existing literature, we
find that central bank communications have an impact on various aspects of the financial
market. \cite{born2014central} concluded his paper that financial markets, arguably is one of the
most important target groups of central bank communications. Their findings suggest that the communications about financial stability have significant outcome on stock prices. \cite{wang2023does} findings focus on dynamics of the economic aspects addressed
in the communications of the Central Economic Work Conference (CEWC) of China and
its connections to the stock market. 

\vspace{\baselineskip}
\noindent
Looking more closely at the literature, we find that central banks' communications have
a significant impact on financial stability. \cite{istrefi2023fed} computed the indicators
that measures the sentiments of the speeches of governors of Fed and FRB presidents.
According to their findings, a higher topic intensity or negative tone, applied to a typical
forward-looking Taylor rule, is linked more to monetary policy accommodation than the
economy’s current position suggests. In an illustration of this, \cite{du2023does} computed
the emotional dictionary of China’s financial stability communications, quantified the tone
of central bank’s financial stability communications, and computed an index for it. The findings demonstrated that: (1) written communications,
in the form of Financial Stability Reports, can successfully steer market trends in the
direction anticipated by the central bank, while oral communications has no discernible
impact; and (2) written communications can lessen stock market volatility, whereas oral
communications can exacerbate stock market volatility. Similarly, \cite{gertler2018central} investigate the impacts of European Central Bank’s communications on the financial market more specifically interest rate, exchange rate, and stock market using least
squares and quantile regressions.

\vspace{\baselineskip}
\noindent
Different textual analysis, natural language processing (NLP), and
deep learning models have been used to quantify the tone of central banks’ communications. Such as \cite{ruman2023comparative} used some methods from (NLP), content analysis, sentiment
analysis, and discourse analysis to understand the decision making process of the Federal Open Market Committee (FOMC) and the factors that impact its monetary policy
decisions especially during high inflation periods. Similarly, \cite{apel2014informative}  via
automatic content analysis, which turns the qualitative information in the minutes of monetary policy meetings into a
quantitative measure, gauged the sentiment and tone of the Swedish Central Bank minutes. They found that this measurement is very useful in forecasting the future policy rate
decisions. To spot the fundamental topics of the central bank speeches, \cite{bohl2023central} used the STM methodology: an unsupervised machine learning model. They inspected
factually how inflation and unemployment expectations impact the tone of the Federal
Reserve System (FED) and the European Central Bank (ECB) relative to past macroeconomic
developments. Moreover, \cite{moniz2014predicting} used textrank link analysis algorithm
and multinomial Naive Bayesian model to predict the effects of Bank of England’s (BoG)
Monetary Policy Committee Minutes on investors’ interest rate expectations. \cite{petropoulos2021can} adopted the analysis of NLP
sentiment index using the XGBoost machine learning technique to identify useful
statements in the corpus of central bank's governors speeches and computed a sentiment index to
predict future financial market behavior. \cite{andersen2023central} applied a multinomial inverse regression (MNIR) to create a positive and negative sentiment dictionary
of the documents of the Norwegian Central Bank. Their thesis tried to investigate the
connection between sentiments in central bank communication and stock market returns.

\noindent
\section{Data}
\subsection{Tone of SBP Communication}
As our paper aims to examine how the tone of the SBP communication influences stock
prices, we require a communication tool with accurate timing. Therefore, we opt for SBP
minutes of meeting since they are posted online on the official website of SBP, ensuring
immediate accessibility for market participants. 

\noindent
The primary challenge in assessing tone involves converting SBP documents into a quantifiable measure suitable for analysis. Initially, we gather documents from SBP officials website accessible online for everyone. The documents presented in real-time are exclusively in the English language, providing a significant advantage for our analysis since
all textual analysis models are trained on English. We perform prepossessing on these documents and convert all the documents into .txt format. We eliminating superfluous elements such as stop words, full stops, commas, punctuation, hashtags, numbers, days, months, years and graphs, and convert all the text to lowercase. The python code for complete prepossessing is in the appendix. We compile a total of 56 documents that span January 2016 to
January 2024.  In particular, these documents published publicly (e.g., the official website of SBP) don't have a regular delivery pattern, as depicted in Fig. 1 below.

\begin{figure}[H]
    \caption{Number of Documents Per Year}
    \centering
    \includegraphics[width=0.8\linewidth]{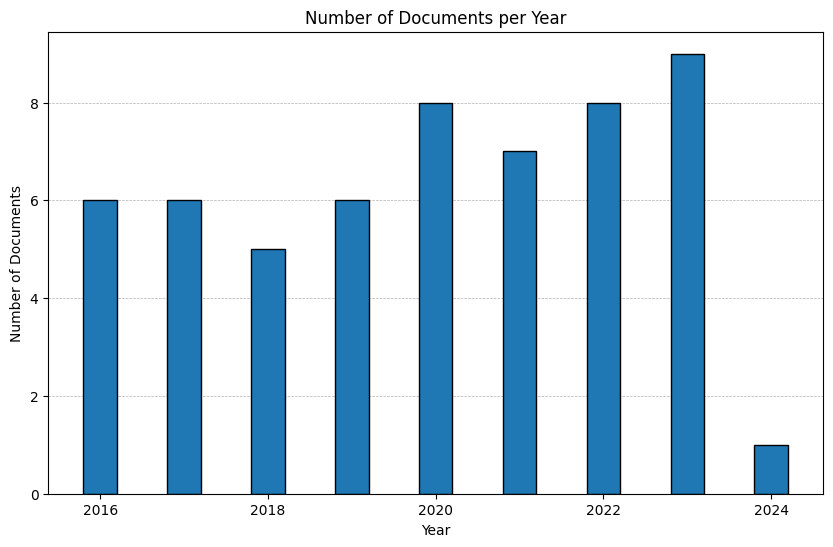}
    \label{fig:1}
\end{figure}

\noindent
In the next phase, we determine the  tone of State Bank of Pakistan (SBP) communications by employing a bag-of-words method and utilizing the financial dictionary created by \cite{loughran2011liability}. This involves analyzing the content of SBP communications using a collection of words associated with financial sentiments from the specified dictionary. We utilize a dictionary method for tone quantification to circumvent  potential drawbacks associated with using the data twice. Specifically, this approach avoids training an algorithm on the same data to which it is subsequently applied. Furthermore, as pointed out by \cite{cambria2017affective}  statistical text classifiers yield meaningful results when provided with a sufficiently large text input \cite{bennani2019does}. To address this, we employ previous information as an LM dictionary to gauge the sentiments of monetary policy documents. This  LM dictionary is specifically constructed for financial as well as economic documents, as explained by \cite{gurun2012don} and \cite{hillert2014media}. This dictionary  has also been proven to be relevant in the context of central bank communication, as discussed by \cite{hansen2016shocking}. \cite{bennani2019does}. Moreover, there are many benefits of the LM dictionary\cite{bennani2019does}. Such as, firstly, researcher subjectivity is avoided \cite{bennani2019does}. Only a relevant package for the LM dictionary has to be installed in Python and run a small chunk of code. A detailed of instructions are available in the appendix.  Ultimately, as this dictionary is accessible to the public, reproducing this analysis is straightforward \cite{bennani2019does}. Numerous studies employ a similar method, using an LM dictionary to tally particular words and sentences to quantify the sentiments in central banks communication, as exemplified in works such as \cite{schmeling2019does} and \cite{hubert2017central}. 

\noindent
That is why, we use the same dictionary technique to find out words which can be classified into positive and negative words in financial context. We count positive words and negative words for each published monetary policy document of State Bank of Pakistan.  

\begin{figure}[H]
    \caption{Percentage of Positive and Negative Words}
    \centering
    \includegraphics[width=0.8\linewidth]{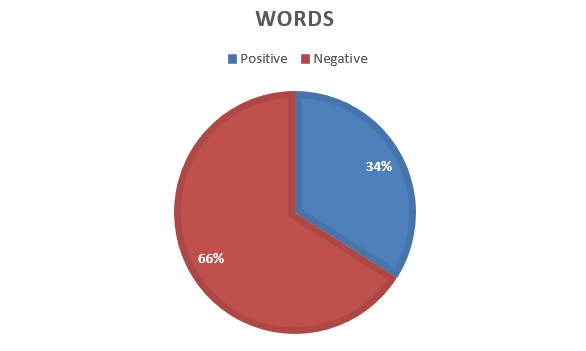}
    \label{fig:2}
\end{figure}

\noindent
Figure 2 shows the total percentage of  positive words and negative words. We can clearly see that SBP documents contain more negative words indicating the pessimism of SBP. 

\noindent
We compute the Tone of SBP documents using the below formula: 
\begin{equation} \label{basic}
\textit{T} = [Positive Words - Negative Words] / 
             [Positive Words + Negative Words]
\end{equation}

\noindent
where \textit{T} represent the tone of  documents. Range of \textit{T} is [-1, +1]. The value close to -1 means the tone of a document is negative, close to zero mean neutral, and close to +1 means positive. 

\noindent
Figure 3 shows the increase and decrease in the tone of SBP monetry policy documents.

\begin{figure}[H]
    \caption{Tone of SBP Communication}
    \centering
    \includegraphics[width=0.8\linewidth]{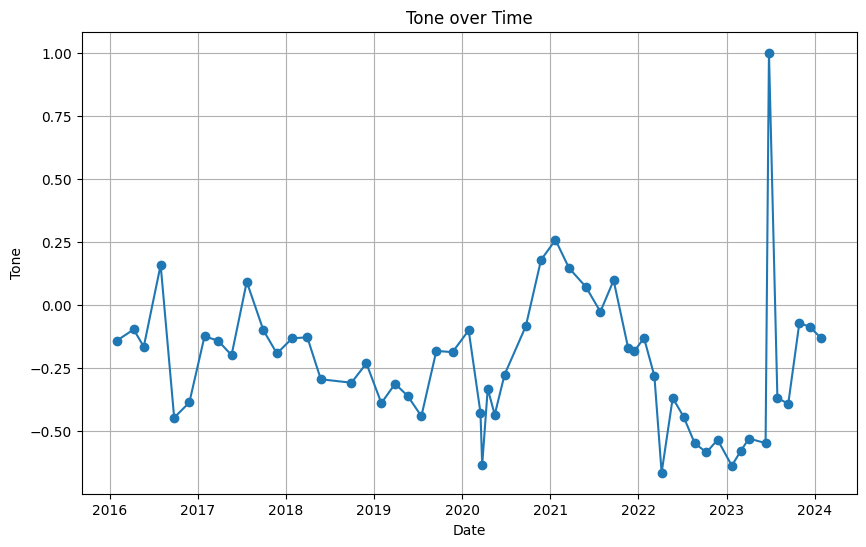}
    \label{fig:3}
\end{figure}

\noindent
 Figure 3 above shows that \textit{T} does not show any apparent trend through the sample period; therefore, it avoids the structural and econometric issues that arise with a trend. The figure 3 also shows that change of the tone evolves negatively and positively through time, suggesting continuous evolution in the tone of SBP communications.

\subsection{ Stock Market Index}
To investigate the influence of the communications of the State Bank of Pakistan (SBP) on the Pakistan Stock Market, we analyze daily stock market data. Our focus is on the KSE - 100 index, which monitors the daily, weekly, and monthly performance of shares from 100 listed companies. Specifically, we examine daily returns within the time frame that coincides with the release of SBP documents. Our goal is to evaluate how the tone of SBP documents released on a given day, influences the stock market's performance on the same day. This analysis employs a high-frequency method to capture the nuanced relationship between SBP communications and the dynamics of the stock market.

\subsection{ Control Variables}

To examine the influence of the tone in State Bank of Pakistan (SBP) communications on stock prices, accounting for economic and financial conditions, we incorporate various control variables. These include data on policy rate, consumer confidence, and economic uncertainty.

\noindent
\textbf{SBP Policy Decision:} Policy rate is one of the main instrument used by central banks for monetary policy. In order to control for SBP policy rate decision, we use monthly KIBOR rate as a proxy. KIBOR rate is calculated monthly.

\noindent
\textbf{Consumer Price Index:} Consumer Price Index (CPI) measures the average change in prices paid by consumers for a basket of goods and services. It is often used to measure inflation. If the inflation is high, it might reduce the purchasing power of consumers, which in return decreases the sales for companies. Change in CPI influences the investors expectations about future economic growth, which can impact their buying and selling in stock market. We use monthly data of CPI because it is measured monthly and yearly. 

\noindent
\textbf{Industrial Production Index:} The Industrial Production Index (IPI) is one of the important factors to gauge the overall health of an economy. Companies in the industrial sector most often have a large share in the stock market. Higher IPI indicates better performance and higher profits for these companies, aids to higher stock prices. We use monthly IPI data.

\noindent
\textbf{Consumer Confidence Index:} Consumer Confidence Index (CCI) measures the overall confidence of consumers and their pessimism and optimism about overall state of the economy and their personal financial conditions. Consumer confidence can significantly impacts the sentiments of investors. When consumers are optimistic, investors feel more confident about future financial conditions, and in return willing to invest more in stock market, and vice versa. We use monthly data of CCI. 

\noindent
\textbf{Economic Uncertainty:} We use the monthly Economic Policy Uncertainty (EPU) index constructed by \cite{choudhary2020measuring} in order to control the political instability and  economic uncertainty in Pakistan. The EPU index is developed using newspapers and articles with respect to policy uncertainty. \cite{choudhary2020measuring} count the number of articles published on the various platform using the terms uncertain or uncertainty, economic or economy, and scale the EPU count by a measure of the number of articles in the same newspaper and month. Higher EPU index decreases the investors' confidence. Investors being uncertain about future policies means they might be reluctant to invest in stocks and vice versa. 

\vspace{1\baselineskip}
\noindent
The sources of our data are Word Bank data base, SBP official website, IMF Data Base, Stock Market data base, and Pakistan Bureau of Statistics. Table 1 below shows the summary statistics of all the variables.

\begin{table}[htbp]
\caption{Summary statistics of variables}
\centering
\begin{tabular}{lrrrrr}
    \toprule
    \textbf{Variable} & \textbf{Obs} & \textbf{Mean} & \textbf{Std. Dev.} & \textbf{Min} & \textbf{Max} \\
    \midrule
    Stock Returns & 56 & -0.001 & 0.014 & -0.068 & 0.032 \\
    Sentiments (Tone) & 56 & -0.224 & 0.279 & -0.666 & 0.999 \\
    Policy Rate & 56 & 11.10 & 5.60 & 5.75 & 22 \\
    Consumer Confidence Index & 56 & 41.90 & 6.70 & 26.20 & 55.9 \\
    Consumer Price Index & 56 & 0.130 & 0.099 & 0.029 & 0.364 \\
    EPU Index & 56 & 128.4 & 62.4 & 28.1 & 328.3 \\
    Industrial Production Index & 56 & -2.80 & 12.5 & -41.1 & 26.8 \\
    KIBOR Rate & 56 & 11.6 & 5.6 & 6 & 23.6 \\
    \bottomrule
\end{tabular}
\label{tab:summary}
\end{table}

\section{Methodology}
In order to check the unit root in our data, to find that whether our data is stationary or not, we use the Augmented Dickey-Fuller (ADF) test. We find that the null hypothesis cannot be rejected for consumer price index, EPU index, and KIBOR rate. Test statistic of all these variables are greater than the critical values at all the levels of significance (1\%, 5\%, and 10\%) as well as their p-values are greater than the common significant level of 0.05. Test statistics of consumer confidence index is less than critical value at 10\% significant level but not than 5\% and 1\% significant levels. But its p-value is greater than the common significant level of 0.05. There is some indication that the consumer confidence index might be stationary, but it's not conclusive. Similarly, The test statistic of industrial production index, which is less than the critical values at the 5\% and 10\% significance levels (but not the 1\% level). This means that we can reject the null hypothesis at the 5\% and 10\% significance levels, but not at the 1\% level. In other words, there is some evidence to suggest that the industrial production index is stationary, but the evidence is not conclusive. Table 2 below shows all the summary of ADF test. 

\begin{table}[H]
\caption{Results of ADF test}
\centering
\begin{tabular}{lrr}
    \toprule
    \textbf{Variable} & \textbf{t-Statistic} & \textbf{Prob} \\
    \midrule
    Stock Returns & -5.44 & 0 \\
    Sentiments (Tone) & -5.35 & 0 \\
    KIBOR Rate & -0.94 & 0.76 \\
    Consumer Confidence Index & -1.98 & 0.29 \\
    Consumer Price Index & -0.63 & 0.85 \\
    EPU Index & -2.69 & 0.08 \\
    Industrial Production Index & -3.24 & 0.02 \\
    \bottomrule
\end{tabular}
\label{tab:adf_results}
\end{table}

\noindent
Since, the data is not completely stationary but composed of both stationary and non stationary variables. Here, the OLS fails to give accurate results. There is a time series component in our data, so there must be the effects of lagged values of the variables. To incorporate such data in our econometrics setup, we used the Autoregressive Distributed Lagged (ARDL) model. This model incorporate lagged values of the dependent variable and possibly the independent variables as well. This accounts for the autocorrelation present in time series data, where current values are often correlated with past values. The ARDL model allows for the estimation of both short-term and long-term effects. The long-term relationship is captured by the coefficients associated with the lagged variables, indicating the equilibrium relationship between the variables \cite{Amal2011ardl}. 

\noindent
The most crucial step after choosing the ARDL model is the number of optimal lags each variable will have. To find the optimial number of lags we use the Akaike Information Criterion (AIC) using E-Views. Rather than concentrating only on the correlation between the time series and its lag values, AIC takes into account both the quality of fit and the complexity of the model. This makes it a superior method \cite{Ahmad2023aic}. Appendix  has the output of AIC.

\noindent
 Figure 4 in the appendix showcases the 20 models with the most favorable AIC values. Each line represents a model, with the y-axis displaying its AIC score and the x-axis differentiating the models. By analyzing the graph, we can determine the model that provides the optimal fit for the data set. In our case, Model3809 emerged as the superior choice, boasting the lowest AIC score on the graph, and gives us optimal number of lags for each specific variable which are 4, 3, 3, 4, 2, 3, 1, also shown in the figure 4.  

 \noindent
\section{Econometrics Setup}
We explore in what ways does a  change in the sentiments of SBP communications impact stock returns. To achieve this, we have to calculate the stock price returns around the day on which SBP monetary policy documents being published. The main motivation behind this is that any price fluctuation in this specific time frame  is probably due to change in the tone of the SBP monetary policy documents. 

\noindent
The State Bank of Pakistan (SBP) encompasses policy targets that extend beyond conventional considerations such as price stability and the output gap to include financial market stability. The release of SBP documents is perceived not just as a communication of policy stance but also as an indication of the State Bank's intent to manage market expectations 
and guide reactions during challenging situations. For instance, a positive change in the tone of these documents might signify the SBP's response to an optimism in the stock market, anticipating government intervention. Consequently, following a negative shock, the stock market can respond negatively to SBP documents. This underscores the role of SBP documents as a policy tool influencing market reactions. To address potential endogeneity issues, we incorporate control variables that could introduce omitted variable bias.

\noindent
To check the instant affect of SBP communications on stock return at month \textit{t} we calculate the daily stock price returns using daily closing price preceding the document and the day on which the document is published.

\begin{equation}
\text{Returns}(t) = \beta_0 + \theta \cdot \text{Tone}(t) + \sum_{i=1}^n \beta_i \cdot X_i(t) + \varepsilon(t)
\end{equation}

\noindent
Returns(t) shows the stock price return where \( t \) showing the publishing day of the SBP monetary policy document, \( \beta_0 \) is a constant, \( \theta \) measures the effect of the change in the tone. \( X(t) \) is the array of control variables with coefficient \( \beta_i \). The error term \( \varepsilon(t) \) contains financial or economic factor affecting the stock returns which have not be taken as control variables.

\section{Results} \label{models}
\begin{table}[htbp]
\caption{Tone of SBP and Stock Market Returns}
    \centering
    \begin{tabular}{lcccc}
        \toprule
        \textbf{Variable} & \textbf{Coefficient} & \textbf{Variable} & \textbf{Coefficient} \\
        \midrule
        Tone        & 0.030*** & KIBOR\_Rate(-2) & 0.009 \\[0.3em]
                    & [0.007]  &                    & [0.004] \\[0.3em]
        Tone(-1)    & -0.010 & KIBOR\_Rate(-3) & -0.010*** \\[0.3em]
                    & [0.007]  &                    & [0.002] \\[0.3em]
        CCI         & -0.003 & CPI & 0.068  \\[0.3em]
                    & [0.005]  &                    & [0.092] \\[0.3em]
        CCI(-1)     & -0.004 & CPI(-1) & -0.109\\[0.3em]
                    & [0.005]  &                    & [0.105] \\[0.3em]
        CCI(-2)     & -0.002 & CPI(-2) & -0.192  \\[0.3em]
                    & [0.006]  &                    & [0.117] \\[0.3em]
        CCI(-3)     & 0.002*** & CPI(-3) & 0.428*** \\[0.3em]
                    & [0.006]  &                    & [0.102] \\[0.3em]
        IPI         & -0.003 & EPU\_Index & 6.60E-05 \\[0.3em]
                    & [0.001]  &                    & [6.15E-05] \\[0.3em]
        IPI(-1)     & 0.009*** & EPU\_Index(-1) & -0.002*** \\[0.3em]
                    & [0.002]  &                    & [5.89E-05] \\[0.3em]
        IPI(-2)     & -0.006 & EPU\_Index(-2) & 0.001 \\[0.3em]
                    & [0.002]  &                    & [6.59E-05] \\[0.3em]
        KIBOR\_Rate & 0.002 & EPU\_Index(-3) & 0.002 \\[0.3em]
                    & [0.002]  &                    & [6.56E-05] \\[0.3em]
        KIBOR\_Rate(-1) & -0.0013 & EPU\_Index(-4) & -6.19E-05 \\[0.3em]
                    & [0.003]  &                    & [4.91E-05] \\[0.3em]
        \midrule
        \textbf{R-squared} & 0.817673 & \textbf{F - Statistic} & 4.312 \\[0.3em]
        \textbf{Adjusted R-squared} & 0.628054 & \textbf{Prob (F-Statistic)} & 0.00 \\[0.3em]
        \bottomrule
    \end{tabular}
    \vspace{1em}
    \par\noindent *, **, *** denote significance at the 10\%, 5\% level, and 1\% level, respectively. Standard errors between brackets.
\end{table}

\noindent
In this section, we explain that there is a statistically significant relation between the sentiments conveyed in the monetary policy documents of the State Bank of Pakistan and stock prices in Pakistan more specifically KSE-100 index returns. To provide evidence that sentiments of SBP monetary policy documents hold further information for stock market fluctuations beyond fundamental economical conditions.  We estimate equation No. 2 twice: without control variables and with control variables. Without control variables results are shown in the appendix. The table 3 below shows the calculated results of equation No. 2 for the time span from 2016M01 to 2024M01.

\noindent
 Below Table  explains  that fluctuation in the sentiments of SBP monetary policy documents has a statistically significant impact on stock prices. Hence, positive (negative) change of the tone is associated with higher (lower) stock prices in KSE-100. The coefficient of tone is significantly positive which is 0.03. In economic terminology, one unit increase (decrease) in tone is associated with an increase of approximately 0.03 units in stock market returns. when we add control variables, the affect of sentiments on stock returns stays the same. More interestingly, with adding lags, the results stays the same as well. Results of base models are in appendix. About control variables, the results shows that consumer confidence index third lag is statistically significant. A unit increase (decrease) in consumer confidence index is connected with an increase (decrease) 0.002 units in stock prices returns. Similarly, the third lag of consumer price index is significant with a coefficient 0.42 indicating that consumer price index has a positive impact on stock returns. The lags of EPU Index and KIBOR rate are also statistically significant with negative coefficients. This result indicates that an increase in either EPU Index or KIBOR rate cause a decrease in stock prices returns. Significance of lag variables indicate that there is not an immediate but delayed affect of control variables on stock market returns.

\noindent
\section{Extension}

\noindent
In this section, we suggest additional extension. We test the constant impact of SBP tone on stock returns. Results shows that there is no constant impact of SBP tone on stock market returns and the impact of tone vanishes day after the monetary policy documents being published.  
\subsection{The Constant Effect of Tone of SBP Communication}
We look into the effect of SBP communication on stock prices days after SBP monetary policy documents being published, publicly. We compute the stock prices over n (number) of days, starting with the publishing day of monetary policy document and denote them as \textit{r}$_{n}$. For instance, when n = 2, \textit{r}$_{n}$ indicates the stock prices return on one day after the monetary policy documents being published. This estimation process have the following equation.

\begin{equation}
r_{n}(t) = \beta_0 + \theta_{n} \cdot \text{Tone}_{n}(t) +  \beta_{i,n}  \sum_{i=1}^{n} \cdot X_{i,n}(t) + \varepsilon_{n}(t)
\end{equation}

\noindent
\textit{r}$_{n}$ is the cumulative stock returns over n days (with n = 2, 3), after the documents of monetary policy published at date t. Rest of the variables are same as equation 2. Table 4 below shows the results for the period 2016M01 to 2024M01. 

\begin{table}[htbp]
\centering
\caption{SBP Tone and Preceding Stock Prices}
\begin{tabular}{cccccc|ccccc}
\toprule
& \multicolumn{5}{c}{Without control variables} & \multicolumn{5}{c}{With control variables} \\
\cmidrule(lr){2-6} \cmidrule(lr){7-11}
Day & $\beta_0$ & $\text{Tone}_{n}(t)$ & $R^2$ & Obs. & & $\beta_0$ &  $\text{Tone}_{n}(t)$ & $R^2$ & Obs. \\
\midrule
$n=2$ & -0.001 & 0.043 & 0.077 & 56 & & -0.016 & 0.019 & 0.751 & 56 \\
& [0.002] & [0.008] & & & & [0.062] & [0.010] & & & \\
$n=3$ & 0.001 & 0.016 & 0.196 & 56 & & -0.056 & 0.003 & 0.725 & 56 \\
& [0.002] & [0.007] & & & & [0.083] & [0.009] & & & \\
\bottomrule
\end{tabular}
\end{table}

\noindent
Table 4 shows the results of impact of the tone on cumulative stock prices. It shows that tone has no significant impact on stock prices during the later time window i.e. when n = 2, 3. Hence, these results suggest that tone of SBP does not have a persistent effect on stock prices a day after the monetary policy documents is published. To save the space, we add the complete results of cumulative stock prices and tone in the appendix.

\section{Robustness} \label{models}
\subsection{Breusch-Pagan-Godfrey test for heteroskedasticity}
\begin{table}[H]
\centering
\caption{Heteroskedasticity Test: Breusch-Pagan-Godfrey}
\begin{tabular}{lrr}
    \toprule
    \textbf{Test Statistic} & \textbf{Value} \\
    \midrule
    F-statistic & 1.68 \\
    Prob. F(26, 25) & 0.09 \\
    Obs*R-squared & 33.0 \\
    Prob. Chi-Square(26) & 0.15 \\
    Scaled explained SS & 12.3 \\
    Prob. Chi-Square(26) & 0.98 \\
    \bottomrule
\end{tabular}
\label{tab:heteroskedasticity}
\end{table}

Table 5 presents the results of the Breusch-Pagan-Godfrey test for heteroskedasticity in our ARDL model. The F-statistic value is 1.683223, with an associated p-value of 0.0986. This indicates some evidence for heteroskedasticity at the 10\% significance level but not at the 5\% level. The Obs*R-squared value is 33.09470 with a p-value of 0.1594, which supports the null hypothesis of homoskedasticity, indicating no significant evidence of heteroskedasticity. Additionally, the Scaled explained SS value is 12.32241 with a p-value of 0.9892, providing very strong evidence for homoskedasticity. Overall, these test statistics suggest that there is no significant evidence of heteroskedasticity in our ARDL model. This implies that the variance of the residuals is constant across observations, ensuring the reliability of our model’s standard errors and test statistics, and thus enhancing the validity of our results. The appendix contains a complete table showing the heteroskedasticity test.

\subsection{Ljung-Box Test for Autorrelation}

\begin{table}[H]
\centering
\caption{Ljung-Box Q-Test Results Summary}
\label{tab:lbqtest}
\begin{tabular}{lrr}
    \toprule
    \textbf{Lag} & \textbf{Q-Statistic} & \textbf{P-Value} \\
    \midrule
    1 & 0.19 & 0.66 \\
    2 & 3.45 & 0.17 \\
    3 & 5.37 & 0.14 \\
    4 & 6.46 & 0.16 \\
    5 & 9.88 & 0.07 \\
    10 & 14.1 & 0.16 \\
    24 & 24.9 & 0.40 \\
    \bottomrule
\end{tabular}
\end{table}

Above table indicates the outcome of the Ljung-Box Q-test for checking for autocorrelation of the residuals of our ARDL model at different lag orders. The table contains three columns: Lag, Q-statistic, and p-value Lag is a measure of how long it takes for the time series to return to a fixed value after a large shock. The first row of the table represents a particular lag and the test statistic (Q-Statistic) and the second row represents the corresponding p-value (P-Value). The Q-Statistic measures the extent of autocorrelation in the residuals at each lag. Higher Q-statistics suggest a greater degree of autocorrelation. The P-Value indicates the probability that the observed autocorrelation could be due to random chance. A low p-value (typically below 0.05) would suggest that the autocorrelation is statistically significant. The Ljung-Box Q-test results in the table indicate that at most lags, the p-values are above the conventional significance level of 0.05. For example, at lag 1, the Q-Statistic is 0.19 with a p-value of 0.66, and at lag 24, the Q-Statistic is 24.9 with a p-value of 0.40. These high p-values suggest that there is no significant autocorrelation in the residuals at these lags. The complete result of the Ljung-Box test is in the appendix. 
\noindent
The lack of significant autocorrelation implies that the residuals are essentially random and not systematically patterned. This indicates that our ARDL model has effectively captured the underlying data patterns and dynamics, leading to reliable and unbiased parameter estimates.

\section{Policy Recommendation} \label{models}
\noindent
To create more stable and predictable financial environment, the SBP should refine its communication strategies with more clear language, more positive words, consistency, and positive reinforcement in its communication regarding monetary policy. Empirical evidence from our study indicates that optimistic and positive sentiments conveyed in monetary policy communications, significantly, give boost to the investors confidence which create more resilient stock market. Therefore, it is recommended that the SBP adopt a proactive communication policy that includes regular updates on economic conditions, transparent explanations of policy decisions, and forward guidance that is both clear and cautiously optimistic. Additionally, leveraging advanced sentiment analysis tools to pre-evaluate the tone of these communications can help ensure that the intended positive sentiment is effectively conveyed. By aligning the tone of its messages with strategic economic goals, the SBP can enhance market predictability, mitigate unnecessary volatility, and support sustained economic growth.

\section{ Conclusion} \label{models}
Using textual, unstructured, and unsupervised data to measure and model the financial dynamics has been very popular since last decade. Our research intends to make contribution to this research by checking the impact of SBP communication on stock market of Pakistan. We first compute the tone of SBP monetary policy announcements by a computational linguistics approach using LM dictionary developed by \cite{loughran2011liability}. In the second phase, we assess the effect of SBP tone on the stock market using a high frequency technique. Results show that improvements in tone have a positive impact on stock prices in Pakistan, particularly in the Karachi Stock Market. Tone has a substantial impact on stock price, and this effect is resistant to a number of control variables. The results of our study demonstrate that SBP communication influences stock price and is important as a tool for policy to influence market expectations. Hence, if Pakistan goes towards flexible regime change and more financial neutrality, transparent communication, might have many advantages to upgrade the impact of State Bank of Pakistan's monetary policy. 

\clearpage
\singlespacing

\clearpage

\section*{Appendix} \label{sec:appendixa}
\addcontentsline{toc}{section}{Appendix}

\noindent

\noindent
The Google Colab notebook in the link contains all the Python code necessary to compute the tone of SBP monetary policy documents. This code provides a step-by-step explanation and performs various tasks, including:

\begin{enumerate}
    \item \textbf{PDF to Text Conversion}: Converts PDF documents to .txt files.
    \item \textbf{Text Preprocessing}: Removes full stops, commas, quotations, dates, days, months, years, hashtags, graphs, and tables from the documents.
    \item \textbf{Sentiment Analysis}:
    \begin{enumerate}
        \item Extracts positive, negative, and neutral words from the monetary policy documents.
        \item Computes subjectivity and polarity scores.
    \end{enumerate}
    \item \textbf{Output}: Saves the results in an Excel file, which includes the number of positive words, negative words, neutral words, subjectivity, and polarity in the monetary policy documents.
\end{enumerate} 
\url{https://colab.research.google.com/drive/1cT6hBM57ZHoOqvdIOgndyIAdvUa9ODeY?usp=sharing}

\vspace{1.5em}
\begin{figure}[H]
    \caption{Optimal Lags Criteria}
    \centering
   \includegraphics[width=0.8\linewidth]{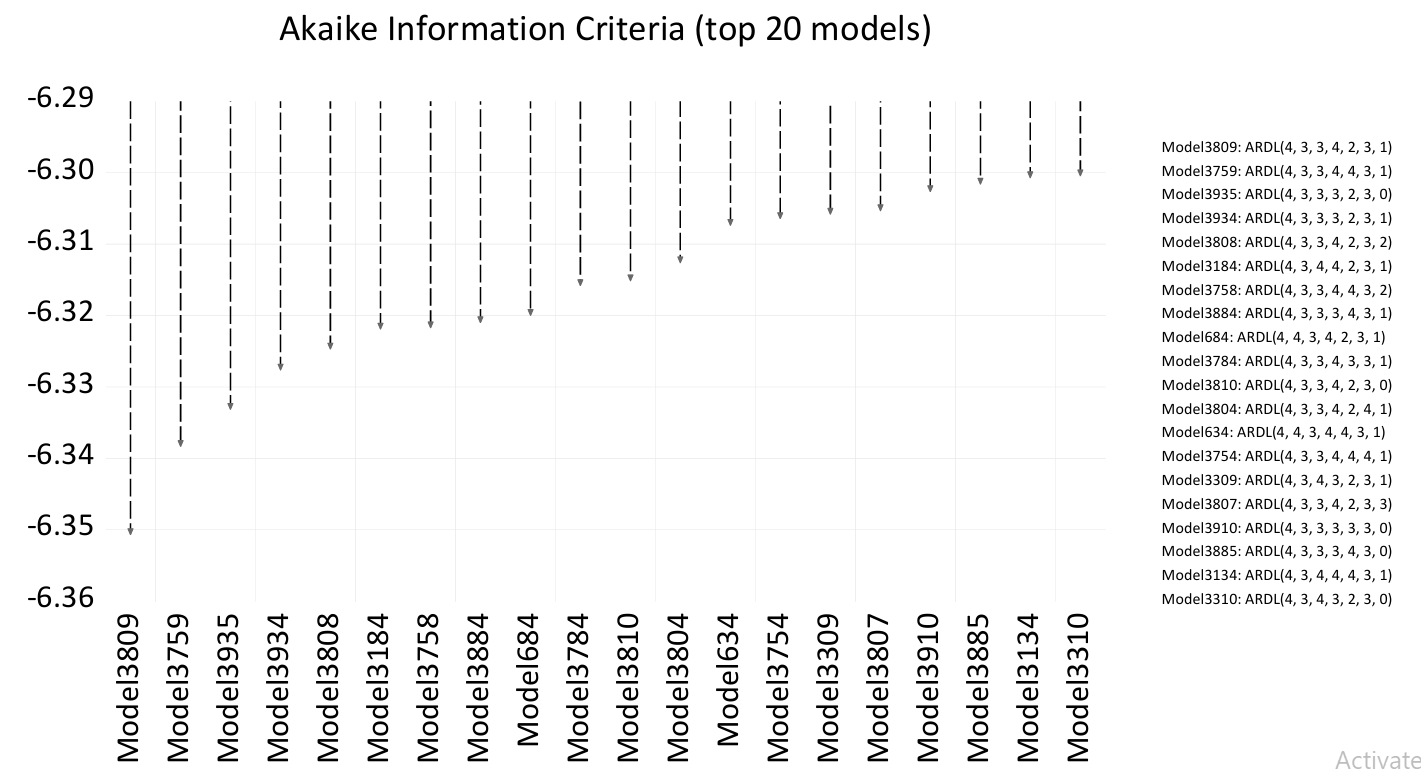}
    \label{fig:1}
\end{figure}

\noindent
The figure 4 above presents the AIC values of the twenty best models estimated from the ARDL approach. On the x-axis, the models are shown starting with Model3809 to Model3310 while the y-axis shows the corresponding AIC values. The AIC values are found to be approximately -6.36 to -6.29, are based on the goodness of fit of the model with lower values considered as better models. Model3809, having the lowest AIC value, is the best model among the top 20 listed models. This model gives optimal lags shown. We incorporate this specific number of lags in our ARDL model.

\vspace{\baselineskip}

\begin{table}[ht]
\caption{Simple OLS Result}
\centering
\begin{tabular}{lcccc}
\toprule
\textbf{Variable} & \textbf{Coefficient} & \textbf{Std. Error} & \textbf{t-Statistic} & \textbf{Prob.} \\
\midrule
C & 0.003 & 0.002 & 1.344 & 0.184 \\
Tone & 0.02** & 0.006 & 2.656 & 0.010 \\
\midrule
R-squared & 0.115 & F-statistic & 7.055 & \\
Adjusted R-squared & 0.099& S.D. dependent var & 0.014 & \\
\bottomrule
\end{tabular}
\label{tab:regression_results}
\end{table}
\noindent
The table 7 above shows the result of OLS model, which has only two variables, tone of SBP monetary policy documents as independent variable and stock market returns. The result indicate that the change in tone has a statistically significant impact on stock prices, even in the absence of lags and control variables.

\vspace{2em}

\begin{table}[H]
\caption{OLS With Lagged Values of Dependent Variable}
    \centering
    \begin{tabular}{lcccc}
        \toprule
        \multicolumn{5}{c}{\textbf{Dependent Variable: Returns}} \\
        \midrule
        Variable & Coefficient & Std. Error & t-Statistic & Prob. \\
        \midrule
        C & 0.002 & 0.002 & 1.215 & 0.230 \\
        Tone & 0.02** & 0.006& 2.605 & 0.012 \\
        Returns(-1) & 0.375** & 0.139 & 2.686 & 0.010 \\
        Returns(-2) & -0.225 & 0.149 & -1.505 & 0.139 \\
        Returns(-3) & -0.095 & 0.148 & -0.643 & 0.523\\
        Returns(-4) & 0.054 & 0.134 & 0.407 & 0.685 \\
        \midrule
        R-squared & 0.274 & F-statistic & 3.488\\
        Adjusted R-squared & 0.196 &  Prob(F-statistic) & 0.009 \\
        \bottomrule
    \end{tabular}
    \vspace{1em}
\par\noindent *, **, *** denote significance at the 10\%, 5\% level, and 1\% level, respectively.
\end{table}

\noindent
Table 8 shows the simple OLS model with only two variables, stock market returns as dependent variable and tone of SBP monetary policy documents as an independent variable alongside four lagged values of dependent variable return. Even in the presence of lagged values of returns, the change in the tone of SBP communication has a statistically significant impact on stock market returns.

\vspace{2em}

\begin{table}[H]
\caption{OLS With Lagged Values of Dependent Variable and Control Variable}
    \centering
    \begin{tabular}{lcccc}
        \toprule
        \multicolumn{5}{c}{\textbf{Dependent Variable: Returns}} \\
        \midrule
        Variable & Coefficient & Std. Error & t-Statistic & Prob. \\
        \midrule
        C & -0.033 & 0.029 & -1.156 & 0.254 \\
        Tone & 0.02** & 0.006 & 2.831 & 0.007 \\
        IPI & 0.002 & 0.002 & 1.130 & 0.276 \\
        Kibor\_Rate & 0.002 & 0.001 & 0.001 & 0.998 \\
        CCI & 0.001 & 0.001 & 0.328 & 0.732 \\
        CPI & 0.048 & 0.062 & 0.785 & 0.433 \\
        EPU & -5.33E-05 & 6.01E-05 & -0.887 & 0.379 \\
        Returns(-1) & 0.152 & 0.203 & 0.780 & 0.428 \\
        Returns(-2) & -0.381 & 0.181 & -2.099 & 0.042 \\
        Returns(-3) & -0.194 & 0.154 & -1.258 & 0.215 \\
        Returns(-4) & -0.063 & 0.154 & -0.411 & 0.682 \\
        \midrule
        R-squared & 0.372 & F-statistic & 2.436\\
        Adjusted R-squared & 0.219 &  Prob(F-statistic) & 0.021 \\
        \bottomrule
    \end{tabular}
    \vspace{1em}
\par\noindent *, **, *** denote significance at the 10\%, 5\% level, and 1\% level, respectively.
\end{table}

\noindent
Table above shows the results of OLS with tone being independent variable along with other control variables and lagged values of dependent variable. The results indicates that the tone of SBP monetary policy documents has a statistically significant impact on the stock market returns. This impact is even stronger than the impact in previous models. 
The interesting fact is that the tone of SBP communication has statistically significant on stock market returns across all the models. After these three base models, we move towards our main ARDL model which is explained in section 6.

\begin{table}[h!]
\caption{Without Control Variables when n=2}
\centering
\begin{tabular}{lcccc}
\toprule
Variable & Coefficient & Std. Error & t-Statistic & Prob. \\
\midrule
Returns\_2(-1) & -0.056 & 0.137 & -0.414 & 0.680 \\
Returns\_2(-2) & -0.273 & 0.138 & -1.972 & 0.054 \\
TONE & -0.0103 & 0.008 & -1.280 & 0.205 \\
C & 0.001 & 0.002 & 0.065 & 0.947 \\
\midrule
R-squared & 0.077 & Mean dependent var & -0.001 \\
Adjusted R-squared & 0.021 & S.D. dependent var & 0.017 \\
F-statistic & 1.396 & Durbin-Watson stat & 2.048 \\
\bottomrule
\end{tabular}
\end{table}

\noindent
Table 10 above is the complete output when we take n=2 (stock returns one day after the documents are published publicly) with tone as an independent variable and no control variables.

\begin{table}[H]
\caption{With Control Variables when n=2}
\centering
\begin{tabular}{lcccc}
\toprule
\textbf{Variable} & \textbf{Coefficient} & \textbf{Std. Error} & \textbf{t-Statistic} & \textbf{Prob.} \\
\midrule
Returns\_2(-1) & -0.432** & 0.166 & -2.601 & 0.015 \\
Returns\_2(-2) & -0.306 & 0.161 & -1.894 & 0.070 \\
Returns\_2(-3) & -0.420** & 0.170 & -2.469 & 0.021\\
TONE & 0.019 & 0.010 & 1.782 & 0.087 \\
TONE(-1) & -0.014 & 0.010 & -1.482 & 0.151 \\
CCI & 0.001 & 0.001 & 1.217 & 0.235 \\
CCI(-1) & -0.003** & 0.001 & -3.389 & 0.002 \\
CCI(-2) & 0.004 & 0.001 & 0.548 & 0.588 \\
CCI(-3) & 0.002 & 0.001 & 0.222 & 0.826 \\
CCI(-4) & 0.001 & 0.001 & 2.004 & 0.056 \\
CPI & 0.149 & 0.136 & 1.093 & 0.284 \\
CPI(-1) & -0.130 & 0.157 & -0.827 & 0.416 \\
CPI(-2) & -0.329 & 0.181 & -1.812 & 0.082 \\
CPI(-3) & 0.118 & 0.153 & 0.776 & 0.445 \\
CPI(-4) & 0.580** & 0.147 & 3.945 & 0.001 \\
EPU & 0.001 & 0.001 & 1.309 & 0.202 \\
EPU(-1) & -0.001 & 0.001 & -0.473 & 0.640 \\
EPU(-2) & -0.001 & 0.001 & -0.669 & 0.509 \\
EPU(-3) & 0.002** & 0.001 & 2.742 & 0.011 \\
IPI & 0.001 & 0.002 & 0.304 & 0.763 \\
IPI(-1) & 0.003 & 0.002 & 1.332 & 0.195 \\
IPI(-2) & -0.003 & 0.002 & -1.287 & 0.210 \\
Kibor\_Rate & -0.003 & 0.002 & -1.050 & 0.304 \\
Kibor\_Rate(-1) & 0.001 & 0.005 & 0.342 & 0.734 \\
Kibor\_Rate(-2) & 0.001 & 0.005 & 0.267 & 0.791 \\
Kibor\_Rate(-3) & 0.003 & 0.005 & 0.753 & 0.458 \\
Kibor\_Rate(-4) & -0.012** & 0.003 & -3.293 & 0.003 \\
C & -0.016 & 0.062 & -0.267 & 0.791 \\
\midrule
R-squared & 0.751 & Mean dependent var & -0.001 & \\
Adjusted R-squared & 0.471 & S.D. dependent var & 0.017 & \\
F-statistic & 2.684 & Durbin-Watson stat & 2.210 & \\
\bottomrule
\end{tabular}
\label{tab:regression_results}
\vspace{1em}
\par\noindent *, **, *** denote significance at the 10\%, 5\% level, and 1\% level, respectively.
\end{table}

 \noindent
Table 11 above is the complete output when we take n=2 (stock returns one day after the documents are published publicly) with tone as an independent variable along with control variables.  

\begin{table}[H]
\caption{Without Control Variables when n=3 }
\centering
\begin{tabular}{lcccc}
\toprule
Variable      & Coefficient & Std. Error & t-Statistic & Prob.* \\
\midrule
Returns\_3(-1)  & 0.162    & 0.132   & 1.066     & 0.249 \\
Tone & 0.016**    & 0.007   & 2.397    & 0.020 \\
Tone(-1)      & -0.012   & 0.007   & -1.684   & 0.098 \\
Tone(-2)      & -0.012   & 0.007   & -1.842   & 0.078 \\
C             & 0.0140    & 0.002   & 5.029    & 0.000 \\
\midrule
R-squared              & 0.196         & Mean dependent var & 0.003 \\
Adjusted R-squared     & 0.131          & S.D. dependent var & 0.014 \\
F-statistic            & 2.998          & Durbin-Watson stat & 2.088 \\
\bottomrule
\end{tabular}
\vspace{1em}
\par\noindent *, **, *** denote significance at the 10\%, 5\% level, and 1\% level, respectively.
\label{tab:regression}
\end{table}

\noindent
Table 12 above is the complete output when we take n=3 (stock returns one day after the documents are published publicly) with tone as an independent variable and no control variables.  

\begin{table}[H]
\caption{With Control Variables when n=3}
\centering
\begin{tabular}{lcccc}
\toprule
Variable          & Coefficient & Std. Error & t-Statistic & Prob.* \\
\midrule
Returns\_3(-1)     & -0.131   & 0.206   & -0.639   & 0.528 \\
Returns\_3(-2)     &  0.229   & 0.185   &  1.231   & 0.230 \\
Returns\_3(-3)     & -0.309   & 0.172   & -1.791   & 0.080 \\
Tone              &  0.003   & 0.009   &  0.331   & 0.742 \\
Tone(-1)         & -0.005   & 0.008   & -0.676   & 0.499 \\
Tone(-2)         & -0.018   & 0.010   & -1.868   & 0.073 \\
Tone(-3)         &  0.011   & 0.008   &  1.340   & 0.179 \\
CCI               & -0.001   & 0.001   & -0.141   & 0.888 \\
CCI(-1)           &  0.001   & 0.001   &  1.508   & 0.144 \\
CCI(-2)           &  0.001   & 0.001   &  1.112   & 0.265 \\
CCI(-3)           & -0.001   & 0.001   & -2.155   & 0.041 \\
CCI(-4)           &  0.001   & 0.001   &  1.079   & 0.281 \\
CPI               &  0.020   & 0.142   &  0.145   & 0.888 \\
CPI(-1)           &  0.106   & 0.168   &  0.629   & 0.529 \\
CPI(-2)           &  0.059   & 0.127   &  0.461   & 0.645 \\
CPI(-3)           & -0.130**  & 0.153   & -0.852   & 0.398 \\
CPI(-4)           &  0.231**   & 0.116   &  1.989   & 0.047 \\
EPU               &  0.001   & 8.35E-05   &  1.463   & 0.144 \\
IPI               & -6.04E-05   & 0.001   & -0.290  & 0.772 \\
IPI(-1)           &  0.002**   & 0.001   &  2.265   & 0.032 \\
IPI(-2)           & -3.59E-05   & 0.001   & -0.143   & 0.886 \\
IPI(-3)           & -0.001   & 0.001   & -0.636   & 0.528 \\
Kibor\_Rate      & -0.003   & 0.002   & -1.366   & 0.171 \\
Kibor\_Rate(-1)  &  0.004   & 0.002   &  1.940   & 0.052 \\
Kibor\_Rate(-2)  &  0.001   & 0.004   &  0.558   & 0.580 \\
Kibor\_Rate(-3)  & -0.004   & 0.005   & -0.963   & 0.345 \\
Kibor\_Rate(-4)  & -0.008***   & 0.003   & -2.964   & 0.004 \\
C                 & -0.056   & 0.083   & -0.674   & 0.506 \\
\midrule
R-squared              & 0.725          & F-statistic            & 2.354 \\
Adjusted R-squared     & 0.417          & Prob(F-statistic)      & 0.018 \\
\bottomrule
\end{tabular}
\label{tab:regression}
\vspace{1em}
\par\noindent *, **, *** denote significance at the 10\%, 5\% level, and 1\% level, respectively.
\end{table}

\noindent
Table 10 above is the complete output when we take n=3 (stock returns one day after the documents are published publicly) with tone as an independent variable along with control variables. 
\end{document}